# Design, Fabrication, and Characterization of a User-Friendly Microfluidic Device for Studying Liver Zonation-on-Chip (ZoC)

Reza Mahdavi[1], Sameereh Hashemi-Najafabadi[2*], Mohammad Adel Ghiass[3], Silmu Valaskivi[4], Hannu Välimäki[4], Joose Kreutzer[4], Charlotte Hamngren Blomqvist[5], Stefano Romeo[6], Pasi Kallio[4], Caroline Beck Adiels[5*]

1 - Biotechnology Department, Faculty of Chemical Engineering, Tarbiat Modares University, Tehran, Iran
2 - Biomedical Engineering Department, Faculty of Chemical Engineering, Tarbiat Modares University, Tehran, Iran
3 - Tissue Engineering Department, Faculty of Medical Sciences, Tarbiat Modares University, Tehran, Iran
4 - Micro- and Nanosystems Research Group, Faculty of Medicine and Health Technology, Tampere University, 33720 Tampere, Finland
5 - Department of Physics, University of Gothenburg, SE-41296 Gothenburg, Sweden
6 - Department of Molecular and Clinical Medicine, Institute of Medicine, Sahlgrenska Academy, Wallenberg Laboratory, University of Gothenburg, Gothenburg, Sweden

**\* Correspond 1:**

Department of Physics, University of Gothenburg, SE-41296 Gothenburg, Sweden
Tel.: +46 31-786 91 23
Tel.: +46 766-22 91 23
E-mail: caroline.adiels@physics.gu.se

**\* Correspond 2:**

Biomedical Engineering Department, Faculty of Chemical Engineering, Tarbiat Modares University, P.O. Box: 14115-114, Tehran, I.R. Iran
Tel.: +98 21-82884384
Fax: +98 21-82884931
E-mail: s.hashemi@modares.ac.ir

**Abstract**

Liver zonation is a fundamental characteristic of hepatocyte spatial heterogeneity, which is challenging to recapitulate in traditional cell cultures. This study presents a novel microfluidic device designed to induce zonation in liver cell cultures by establishing an oxygen gradient using standard laboratory gases. The device consists of two layers; a bottom layer containing a gas channel network that delivers high and low oxygenated gases to create three distinct zones within the cell culture chamber in the layer above. Computational simulations and ratiometric oxygen sensing were employed to validate the oxygen gradient, demonstrating that stable oxygen levels were achieved within two hours. Liver zonation was confirmed using immunofluorescence staining, which showed zonated albumin production in HepG2 cells directly correlating with oxygen levels and mimicking in-vivo zonation behavior. This user-friendly device supports studies on liver zonation and related metabolic disease mechanisms in vitro. It can also be utilized for experiments that necessitate precise gas concentration gradients, such as hypoxia-related research areas focused on angiogenesis and cancer development.

# 1 Introduction

The liver is composed of a set of highly dedicated cell types which organize into functional units known as lobules. Here, the cells specialize in essential tasks such as detoxification, bile production, protein synthesis, glucose regulation, and vitamin storage (Arias et al., 2020). Within each lobule, multiple blood capillaries, called sinusoids are arranged radially, forming an acinus that culminates at the central vein (Juza & Pauli, 2014), as depicted in Figure 1. At the periportal end of the lobule, the hepatic artery and portal vein converge to supply the cells with oxygen and nutrient-rich blood. As cells consume these resources, concentration gradients form along the sinusoids, with the highest concentration near the periportal region and the lowest near the central vein at the pericentral end. Consequently, cells encounter varying concentrations of nutrients, oxygen, and waste products based on their location, driving the phenomena of metabolic zonation within the liver lobule (Tomlinson et al., 2019). Metabolic zonation is a critical physiological feature that allows liver cells to exhibit diverse activities and characteristics depending on their position along the sinusoidal space. This zonation is often overlooked in conventional cell cultures, raising concerns about the accuracy of *in-vitro* models (Scheidecker et al., 2020).

The emergence of microfluidic technology has spurred efforts to develop *in-vitro* liver models that replicate the intricate microstructural functions found *in-vivo* (Leung et al., 2022). Microfluidics offers advantages such as precise spatiotemporal control over cell cultures by adjusting parameters like shear stress and availability of oxygen and nutrients, enabling cells to experience physiologically relevant environmental conditions (Rashidi, Alhaque, Szkolnicka, Flint, & Hay, 2016; Vinci et al., 2011). Hence, microfluidics-based *in-vitro* cultures hold significant promise as tools for assessing drug efficacy and toxicity, effectively bridging the gap between preclinical and clinical testing (Materne, Tonevitsky, & Marx, 2013; Zhang, Korolj, Lai, & Radisic, 2018).

Incorporating zonation into an *in-vitro* liver model allows for the examination of specific niches in the context of disease or metabolic disruptions. For instance, hepatic steatosis predominantly affects hepatocytes in the pericentral zone, where the fatty acid beta-oxidation takes place (Brunt, 2010). Conversely, hepatitis resulting from a yellow fever viral infection selectively spares hepatocytes around the periportal zone (Quaresma et al., 2006). Therefore, liver models with zonation features facilitate the investigation of liver diseases within the specific zone at the very location in which they are manifested. Additionally, these models allow for the exploration of inter-zonal communication between cells, providing insights into the mechanisms underlying liver diseases and metabolic disorders (Bale et al., 2014).

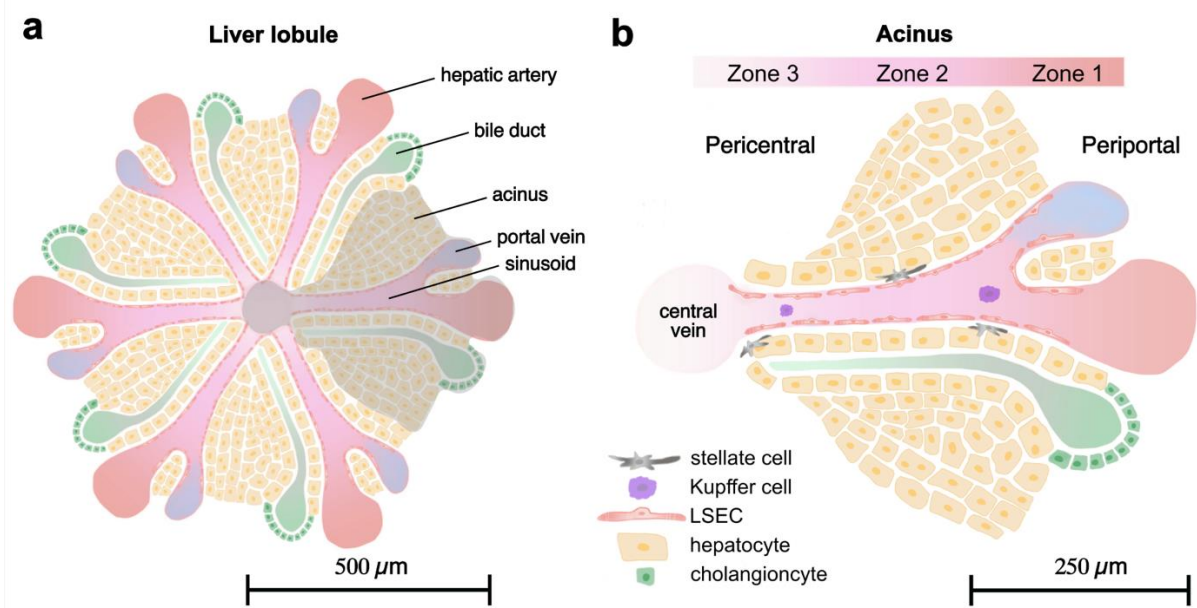

**Fig. 1** Schematic of the liver lobule and the influence of sinusoidal structure on hepatic zonation. (a) The liver lobule consists of hexagonal structures housing predominantly hepatocytes. The blood is supplied by the periportal vein and hepatic artery, providing nutrient and oxygen-rich blood respectively, which converge into the liver sinusoids. (b) The acinus is structured into three zones based on the hepatic function and metabolic activity. Endothelial cells (LSECs) line the vascular wall, shielding hepatocytes from shear stress while allowing constituents from the blood to diffuse into the tissue. As blood moves from the periportal side of the sinusoid toward the central vein, cellular consumption creates gradients of nutrients and oxygen along the sinusoid. These gradients are crucial in driving hepatic zonation, where hepatocytes display distinct functional behaviors depending on their location within the hepatic lobule. Additionally, other cell types such as stellate cells and liver-specific macrophages, known as Kupffer cells, each fulfill specific roles in both healthy and diseased liver tissue

Different biomolecules display gradients within the liver acinus, where the concentration of dissolved oxygen plays a critical role in inducing zonation through β-catenin signaling (Kietzmann, 2017; Tomlinson et al., 2019). In laboratory settings, scientists employ microfluidic liver-on-a-chip platforms to establish oxygen concentration gradients in cell cultures. Various methods, such as passive induction, the use of chemical oxygen scavengers, and the introduction of gaseous oxygen, are employed to generate gas gradients in these devices (Ghafoory et al., 2022; Y. B. Kang, Eo, Bulutoglu, Yarmush, & Usta, 2020; Y. B. A. Kang, Eo, Mert, Yarmush, & Usta, 2018). However, each method comes with its own inherent limitations. When employing passive induction, the generated gradient of nutrients and oxygen in the cell culture medium is solely driven by cellular metabolism, reflecting the natural zonation principles observed in liver-on-a-chip studies (Domansky et al., 2010; Lee-Montiel et al., 2017; Li, George, Vernetti, Gough, & Taylor, 2018). Nevertheless, the sensitivity of oxygen concentration to changes in flow rate poses challenges to controlling the microenvironment, and fluctuating cellular metabolism contributes to the issue. Adjusting the flow rate to modify the gradient may expose cells to varying shear stresses, potentially eliciting unwanted or detrimental cellular responses. Thus, the range of flow rate adjustments is limited and depends on cell type and microfluidic channel dimensions. Another approach to modulate the dissolved oxygen concentration in a microfluidic chip could be achieved by adding chemical oxygen scavengers such as sodium sulfate, cobalt nitrate, or deferoxamine (DFX) to the cell culture medium (Ghafoory et al., 2022; Y. B. Kang et al., 2020; Y. B. A. Kang et al., 2018). One major limitation with this approach is that scavengers alter the electrolyte balance of the cell culture medium and can adversely affect cells (Han et al., 2020). A more cell friendly option is to introduce gaseous oxygen through a permeable substrate like a polydimethylsiloxane (PDMS) membrane (Tonon et al., 2019; Tornberg et al., 2022). Typically, this non-invasive method necessitates access to gas facilities and large containers of costly oxygen mixtures, which are commonly available only in highly specialized mammalian cell culture laboratories.

In this study, we present a novel and user-friendly microfluidic device designed for 2D liver zonation-on-a-chip (ZoC) applications. The ZoC device induces zonation by adjusting the oxygen concentration available to the cultured cells. To regulate the oxygen levels in the ZoC culture chamber, we utilize ambient incubator air and pure nitrogen as gas sources, both of which are readily accessible in standard laboratories. This approach not only allows for precise control of oxygen tension but also simplifies operational procedures for users. The device comprises a culture chamber where cells are housed, with oxygen diffusing from a network of gas channels underneath. This setup creates three distinct oxygen tension zones within the chamber, closely mimicking the *in vivo* environment. Concurrently, cells in these zones receive cell culture medium sequentially, replicating the nutrient gradient observed *in vivo* within the acinus. The microfluidic design, along with the distribution of glucose, oxygen, and shear stress, was simulated using COMSOL Multiphysics, both with and without cells in the ZoC device's culture chamber. Experimental measurements using luminescence-based ratiometric oxygen sensors verified the oxygen tension levels. Finally, the viability of HepG2 cells was confirmed using the device, evaluating functionality based on spatially zonated albumin expression.

## 2 Materials and methods

## 2.1 Device design and fabrication

The device is fabricated in polydimethylsiloxane (PDMS) and comprises a dual-layer system designed to accommodate a 2D layer of perfused cells subjected to different oxygen levels. The lower layer (gas channel network) contains two non-intersecting gas channels measuring 0.2 mm in width and approximate 2 mm in thickness. The upper layer (cell chamber) has a thickness of approximately 3 mm. Together, these layers contribute to an overall device thickness of about 5 mm bonded to a cover slide (see Figure 2).

The gas channel network is supplied with two different gases: a typical high-oxygen cell incubator supply gas (19% oxygen) and pure nitrogen gas (0% oxygen). This lower layer is plasma bonded facing the cover slide, leaving the gas to permeate through the PDMS material to reach the cells (see Figure 2a). To achieve zonation, a suction is applied at one outlet to guide the ambient air from the cell incubator into the chip at a flow rate of 2-3 mL/min, while nitrogen is introduced into the gas channel from a bottle at the opposite side at an equal flow rate (see Figure 2b). The design of the gas channel network has been carefully engineered to generate three distinct oxygen levels within the cell culture chamber, mirroring the *in-vivo* environment.

Initially, the cell chamber design and the corresponding photomask for the gas channel network pattern was generated using Corel DRAW. The gas layer was fabricated in two steps using soft lithography according to the methodology by Banaeiyan et al. (Banaeiyan et al., 2017). In short, a layer of the anti-photoresist SU-8 3050 was spin-coated for 30 s at 1000 rpm onto a silicon wafer, leaving a layer of 100 μm thickness of SU-8. The photomask was then placed on top the SU-8 coated wafer and the entire wafer was exposed to UV radiation for one minute (MA200 mask aligner, SUSS Microtech). The wafer was then developed using a developer solution (Micro Resist Technology GmbH), to remove the unexposed photoresist, thereby revealing the patterned mold.

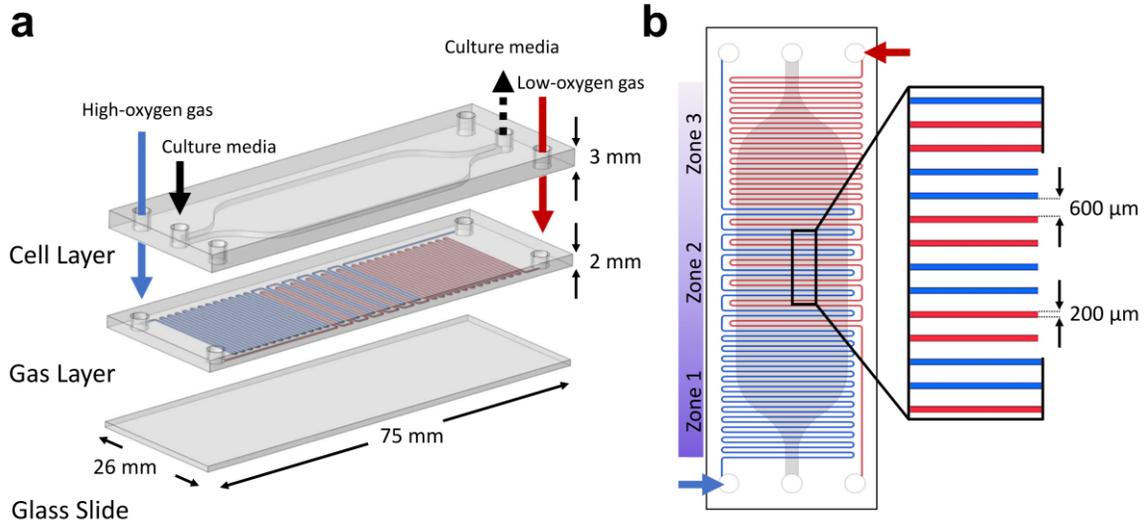

**Fig. 2** The computer-assisted design (CAD) of the chip layers and detailed gas channel network. **a** The split view, which consists of the upper cell chamber layer hosting the cells featuring an inlet (the solid black arrow) and an outlet (the dotted black arrow), a gas channel network layer of serpentine gas channels in the middle (the inlets for high-oxygen gas and low-oxygen gas are shown in blue and red arrows, respectively), and a glass slide at the very bottom. **b** The top view of the device, displaying the overlap of the gas and cell layers. It illustrates the positions of the gas channel inlet and outlets, demonstrating how these channels establish three distinct oxygenation zones within the cell chamber: high (zone 1, corresponding to the periportal zone), medium (zone 2) and low (zone 3, corresponding to the pericentral zone)

To create the cell chamber mold, a 1 mm sheet of poly(methyl methacrylate) (PMMA) was laser-cut based on the computer-assisted design (CAD) software design. The resulting structure was affixed to a flat glass surface using commercially available adhesive super glue (Henkel, Düsseldorf, Germany).

For the soft lithography procedure, a mixture of PDMS (Sylgard 184, Dow Corning, Midland, Michigan, US) and its corresponding crosslinker was prepared in a 10:1 (w/w) ratio. After degassing, this mixture was poured onto both the cell and gas layer molds, and left to incubate at 60°C overnight for crosslinking. After carefully removing resulting crosslinked PDMS from the molds, inlet and outlet holes were created using a 1.5 mm puncher (33-31AA-P/25, Miltex, Integra LifeSciences, Princeton, New Jersey, US). The two layers were precisely aligned and bonded together via air plasma treatment (18 W, 30 s; PDC-32G-2, Harrick Plasma, Ithaca, New York, US), followed by incubation at 60°C for 30 minutes. Subsequently, the device was plasma bonded onto a plain soda-lime glass slide (76 × 26 × 1 mm from Paul Marienfeld GmbH & Co. KG, Lauda-Königshofen, Germany) and incubated overnight at 60˚C before use.

## 2.2 Numerical analysis

The numerical analysis was performed in COMSOL Multiphysics 6.0 finite element software (Burlington, MA, US) as a three-dimensional domain. The simulation model incorporated a coupled analysis of computational fluid dynamics (CFD) within the culture chamber and simultaneous glucose and oxygen mass transfers. While fluid flow and glucose distribution were confined to the culture chamber, the analysis of oxygen involved the entire device, including the PDMS bulk and gas network.

The flow simulation assumed Newtonian incompressible laminar flow with a flow rate of 0.5 µL/min and no-slip boundary conditions at the channel walls. The outlet pressure of the device was set to atmospheric pressure. Density and dynamic viscosity values were assigned as 1.009 g/cm$^3$ and 0.93 mPas, respectively, based on the properties of the DMEM culture media supplemented with 10% FBS at 37°C (Poon, 2020).

The fluidic behavior of the flow in the laminar regime was simulated under the assumptions of constant fluid density and viscosity throughout the culture chamber, and it coupled the mass conservation and Navier-Stokes as the governing equations:

$$\rho \nabla \cdot u = 0,$$

and

$$\rho \left( \frac{\partial u}{\partial t} + u \nabla u \right) = -\nabla p + \mu \nabla^2 u + F,$$

where $\rho$ is the density of the fluid, $u$ is the velocity vector, $t$ is time, $p$ is the pressure, $\mu$ is the dynamic viscosity, $\nabla$ is the gradient operator, $\nabla^2$ is the Laplacian operator, and $F$ is any external force acting on the fluid.

We utilized the convection-diffusion equation to simulate the mass distribution of oxygen and glucose:

$$\partial C / \partial t + \nabla \cdot (uC) = D \nabla^2 C.$$

Here, $C$ is the concentration of the species being transported, $t$ is time, $u$ is the fluid velocity, and $D$ is the diffusion coefficient set to $6.16 \times 10^{-10}$ m²/s and $2.69 \times 10^{-9}$ m²/s, for glucose and oxygen in the cell culture medium at 37°C, respectively (Bavli et al., 2016; Place, Domann, & Case, 2017).

For modeling the oxygen distribution, the oxygen diffusion coefficient in PDMS was set to $3.25 \times 10^{-9}$ m²/s (Markov, Lillie, Garbett, & McCawley, 2014), with saturated oxygen concentration applied at the boundaries exposed to the atmosphere, measuring 1.8 mol/m³. The interface between the cell culture medium and PDMS was characterized using a partition coefficient of 10 (Shiku et al., 2006), and the cellular oxygen consumption rate (OCR) was modeled using the Michaelis-Menten equation:

$$OCR = \frac{qN_t}{A_t} \left( \frac{C}{(K_m + C)} \right),$$

where $K_m$ is the Michaelis-Menten constant equal $6.3 \times 10^{-3}$ mol/m³ (Wagner, Venkataraman, & Buettner, 2011), $q$ is the OCR for each cell, $N_t$ is the total number of cells, $A_t$ is the cell chamber area, and $C$ is the local concentration of oxygen at the lower surface of the culture chamber. The cellular OCR, denoted as $q$, was set to match that of hepatocytes, specifically $3.5 \times 10^{-16}$ mol/s (Wagner et al., 2011). Also, a Heaviside function was used to account for cell necrosis resulting from hypoxia. The critical oxygen concentration at which the cells undergo necrosis was defined as $1.0 \times 10^{-4}$ mol/m³ (Buchwald, 2009). For oxygen concentration simulation in the whole device, two scenarios were considered: one with cells present (400k cells/mL of seeding density) and another without cells. In each scenario, both static and time-dependent modes were employed to analyze the equilibrium and dynamic behavior of the system, respectively. A flow rate of 0.5 µL/min was selected for both scenarios.

At the cell chamber bottom surface, a negative flux was introduced into the system based on the calculated glucose consumption rate of the HepG2 cell line. The simulation of glucose distribution was performed in conjunction with the oxygen distribution study within the cell chamber. Initially, glucose concentration at the entrance of the cell chamber was set to the glucose content of the culture media at 5.5 mM. To estimate glucose consumption, we utilized the glucose uptake rate for the hepatic cell line, which is $2.4 \times 10^{-9}$ mol/min per million cells (Bavli et al., 2016). This rate was then normalized based on the total cell count and culture area within the device.

## 2.3 Ratiometric oxygen sensing

To analyze the microfluidic spatiotemporal oxygen pattern, a ratiometric 2D oxygen imaging system was utilized, following the methodology outlined by Ungerböck et al. (Ungerböck, Charwat, Ertl, & Mayr, 2013). This system utilizes the red and green channels of an RGB camera as the oxygen-sensitive, and reference channel, respectively. Oxygen levels were measured using Platinum (II)-5,10,15,20-tetrakis-(2,3,4,5,6-pentafuorphenyl)-porphyrin (PtTFPP) (Livchem Logistics GmbH, Frankfurt, Germany) as the oxygen-sensitive dye and Macrolex Fluorescent Yellow (MFY) (Livchem Logistics GmbH, Frankfurt, Germany) as the reference dye. Oxygen sensing films with a thickness of approximately 14 μm were knife-coated on glass plates, as described in detail by Tornberg et al. (Tornberg et al., 2022).

A CCD RGB camera (Axiocam 503) mounted on an inverted fluorescence microscope (Zeiss Axio Observer Z1) was used for the oxygen imaging. Images were acquired with a 2.5X objective (2.5x/0.06) and using an FITC filter set consisting of a bandpass 450–490 nm excitation filter, a 510 nm dichroic mirror and a long pass 515 nm emission filter. A 455 nm LED source (M455L4-C4, 690 mW) was used for the illumination.

The sensor material was blended with polystyrene (PS) that has a substantially lower oxygen permeability compared to PDMS, and hence could impair the oxygen diffusion within the device. Therefore, instead of covering the entire cell culture surface, small sensor sections were incorporated on the bottom of the cell chamber. These sensor sections, each measuring 1.5 mm in diameter, were created by first detaching the knife-coated film from the glass plate by immersing the oxygen sensing plate in deionized water, and then placing the detached sensor film on a PDMS membrane, and finally cutting circular pieces from a sensor film using a biopsy puncher. This approach not only ensures the effective functioning of the device gas exchange with real-time oxygen sensing, but also allows for microscopy imaging of the remaining cell culture areas, as the rest of the cell culture surface is optically transparent [28]. During the device fabrication process, ten sensor film sections were gently bonded to the bottom of the culture chamber using a plasma bonding. To ensure homogeneity for cell attachment, this surface received a thin layer of spin-coated PDMS (approximately 10 μm) using 0.5 mL PDMS at 1000 rpm for three minutes. The fabrication process was finalized by incubating the device at 70°C for 15 minutes to polymerize the spin-coated PDMS.

## 2.4 Oxygen measurement and calibration

Prior to the measurement, the high-oxygen gas (19% oxygen) was infused into both gas channels for a minimum of one hour to establish equilibrium within the device. Subsequently, one of the gas channels was switched to 0% oxygen (i.e., nitrogen gas), while the other continued to receive a continuous flow of high-oxygen gas. Imaging started immediately, capturing fluorescence images at 10-minute intervals. At the end of the experiment, a sodium sulfite solution (Sigma Aldrich, Missouri, USA) with a concentration of 10 mg/mL was injected into the cell chamber to scavenge oxygen and ensure 0% oxygen concentration for calibration purposes.

For image analysis, a customized MATLAB program was applied. Within each defined region of interest (ROI), we computed the ratiometric parameter value $R$ by dividing the red channel mean value by the green channel mean value. Subsequently, the ratiometric parameter values for each gas mixture were subjected to fitting using the simplified two-site Stern–Volmer equation (Demas, DeGraff, & Xu, 1995):

$$\frac{R}{R_0} = \frac{f_1}{1 + K_{sv}[O_2]} + f_2 \,.$$

Here, $R$ represents the ratio of the dye emissions at an oxygen concentration of $[O_2]$, and $R_0$ represents the ratio of the dye emissions in the absence of oxygen. Additionally, $f_1$ represents the fraction of dye molecules that are available to quenching and is associated with the Stern-Volmer coefficient $K_{sv}$, while $f_2 = 1 - f_1$ represents the fraction of dye molecules that cannot be quenched. The fitting procedure provided estimations for $f_1$ and $K_{sv}$, which were subsequently used to determine the oxygen concentration during the measurements. Calibration procedures were conducted for each ROI prior and after each measurement for 19% and 0% oxygen, respectively.

## 2.5 Cell culture

HepG2 cells (ATCC, Menassas, VA, USA) were cultured in T-75 flasks using Minimum Essential Media/Earle's Balanced Salt Solution (MEM/EBSS) medium (SH30244.01, HyClone, GE Healthcare, Chicago, Illinois, USA) supplemented with 10% Fetal Bovine Serum (FBS) (SV30160.03, HyClone, Logan, UT, USA), 1% penicillin-streptomycin (SV30010, HyClone), 1% L-glutamine, 1% sodium pyruvate (BE13-115E, Lonza, Basel, Switzerland), and 1% non-essential amino acids (SH30238.01, HyClone). The cells were maintained in a humidified incubator at 37°C with 5% $CO_2$ until they reached a confluency of 70-80%. Subsequently, the cells were detached from the flasks using TrypLE Express (Gibco, Thermo Fisher Scientific, Waltham, MA, USA), centrifuged at 1000 rpm for 2 minutes and the supernatant was discarded. The cells were resuspended in cell culture medium to achieve a final concentration of $4 \times 10^5$ cells/mL. Cells used for culturing in the device did not exceed passage number 25.

To ensure sterility, the devices were sterilized using 70% ethanol, followed by rinsing with sterile de-ionized water. The cell chamber was coated with poly-Lysine and incubated at room temperature for two hours. Subsequently, a HepG2 cell suspension (400k cells/mL) was introduced into the device via a micropipette and allowed to attach for two hours. The device was then connected to a syringe pump (CMA 400, CMA Microdialysis, Kista, Sweden) set at a flow rate of 0.5 μL/min. After connecting the gas lines to the device, it was placed in a humidified incubator for 24 hours at 19% $O_2$, 5% $CO_2$ and 37°C.

## 2.6 Cell viability and imaging

To assess the viability of HepG2 cells within our microfluidic device, we utilized a live/dead staining assay (L3224; Thermo Fisher Scientific). This assay utilized a mixture of calcein-AM (1:1000 v/v), ethidium homodimer-1 (1:1000 v/v), and Hoechst (1:2000 v/v) in phosphate-buffered saline (PBS). These stains labeled living cells, dead cells, and total cell nuclei, respectively. The cells were exposed to the staining solution for 15 minutes at 37°C in a light-protected environment and then rinsed with PBS containing calcium and magnesium. Imaging was carried out using a Leica fluorescence microscope (DMI6000B microscope, Leica Microsystems, Wetzlar, Germany). For quantitative analysis, the acquired images were processed using ImageJ software. After applying the appropriate threshold, the numbers of living, dead, and total cells were determined based on the green, red, and blue channels, respectively. The percentage of living cells was calculated using the following formula:

$$Viability\ (\%) = \frac{total\ cells - dead\ cells}{total\ cells} \times 100.$$

## 2.7 Albumin immunofluorescence staining

After a 24-hour exposure to the oxygen gradient, the microfluidic device was disconnected from the syringe pump. To access the cell chamber area, the cell chamber was opened from the top by cutting around the edges of the chamber walls with a scalpel. A washing procedure was then carried out using PBS at 37°C. Subsequently, the cells were fixed using 4% formaldehyde for 15 minutes at room temperature, followed by three consecutive washes with PBS. The cells were then permeabilized with 0.25% Triton X-100 in PBS for five minutes, followed by three additional PBS washes. To block non-specific antibody binding, a solution of dried milk (5 mg/mL) was applied. The cells were then incubated overnight at 4°C with a diluted primary antibody solution specific to albumin (1:200 goat anti-albumin, Bethyl Laboratories, A80-129A) in PBS. The following day the secondary antibody (1:2000 Donkey anti-goat, Alexa Fluor 594, Thermo Fisher Scientific, A-11058) diluted in 1% BSA was added to the samples. After two hours at room temperature in the dark, the samples were washed three times with PBS. To counterstain the cell nuclei, Hoechst (1:2000, 33258; Thermo Fisher Scientific) was applied for 15 minutes, and the cells were subsequently washed with PBS. The imaging process was performed as previously detailed in the earlier section.

## 2.8 Quantification of immunofluorescence staining

Automated fluorescence imaging involved capturing 36 images of each zone, arranged in a square array at 20x magnification. Prior to image acquisition, meticulous positioning and focusing were applied to each imaging site. To avoid image overlap of the readouts and minimize the risk of nearby imaging sites being exposed to excitation light, which might lead to bleaching and impact results, an offset equal to 50% of the field of view size was chosen between each image. The image scanning started from one corner, moving back and forth to cover all the designated imaging sites.

Images were saved in the raw format to facilitate subsequent image channel separation through Leica Application Suite X (LAS X) software. For immunofluorescence staining quantification, each image was converted to 8-bit and grayscale using ImageJ, with the greyscale value of each fluorescent channel extracted. To normalize the signal, the red signal value (indicative of albumin staining) was divided by the blue signal value (representing Hoechst staining of the cell nuclei) and presented as relative fluorescent units (RFU) for comparative analysis.

## 2.9 Statistical Analysis

The results were derived from a minimum of three independent experiments (n=3), and values are presented as the mean ± standard deviation (SD). The statistical method employed for hypothesis testing was a one-way Analysis of Variance (ANOVA), with a significance level set at $p < 0.05$ to denote statistical significance.

# 3 Results

## 3.1 Numerical analysis of device performance

The numerical analysis of the device performance was focused on the zonation regarding the distribution of glucose and oxygen, as well as on the shear stress within the device. Since the presence of cells contribute to the zonation, the CFD simulations were performed both with cells and in a cell-free scenario. At a flow rate of 0.5 µL/min and a seeding density of 400k cells/mL, most of the culture area experienced shear stress around $2.8 \times 10^{-6}$ Pa. The exception was near the chamber walls where the shear stress was lower. Figure 3a shows the simulated shear stress profile along a line perpendicular to the culture surface of the ZoC. At this flow rate, the simulated glucose concentration decreased slightly from 5.5 mol/m$^3$ at the inlet to 4.2 mol/m$^3$ at the outlet when considering cellular metabolism (Figure 3b). Oxygen measurement simulations were conducted using numerical models that incorporated 2D circular structures at the cell chamber bottom surface. These structures were designed to resemble the sensor section locations for ratiometric oxygen measurement (see Section 1.3 for details). Figure 3d illustrates simulations of the three levels of oxygen concentration at equilibrium ranging from 19% at the media inlet, to 7% at the outlet, and approximately 13% in-between. Each curve corresponds to a circular structure mimicking an oxygen sensor. These circular structures serve as regions of interest (ROIs) for dynamic oxygen measurement simulations, see Figure 3d-e. For each time step during the simulation, the average oxygen concentration in each ROI was plotted against time.

Figures 3d and 3e show the dynamics of the oxygen concentration throughout the simulations, comparing scenarios of with and without cells present. At the beginning of the simulations, we introduced high (19%) and low (0%) oxygenated gases into the initially oxygen-saturated (19%) gas channel of the device. Regardless of whether the simulations accounted for the cells in the chamber or not, oxygen concentrations quickly diverged and reached distinctly separated zones already within a few minutes, stabilizing around 30 minutes later. Importantly, when including cells in the simulations, oxygen concentrations were notably lower, particularly in zone 3 where the concentrations were the lowest.

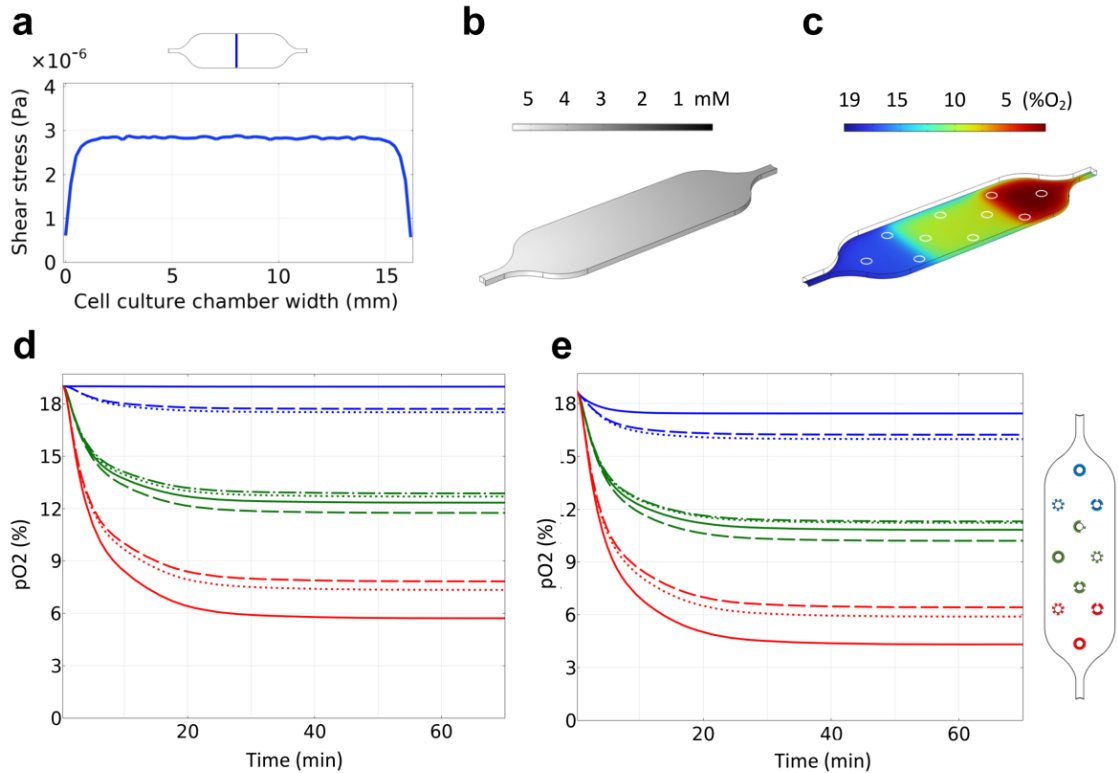

**Fig. 3** Simulation results of the ZoC device. Panel **a** shows shear stress magnitude along the width of the culture area. Panel **b** shows the glucose distribution in the device considering the consumption by HepG2 cells at a density of 400k cells/mL, while panel **c** illustrates the local oxygen concentration at the culture (bottom) surface in the cell chamber and the locations of the regions of interest (ROIs). Both panels depict simulations where high-oxygenated gas is introduced at the lower end of the device schematic. **d** The data from each ROI (as shown in panels **d** and **e**) within the chamber was recorded after simulating the initiation of the gas flow, depicting the oxygen dynamics within the device under two conditions; one without considering the effect of cells **d** and the other when considering the effect of cells **e**. The blue, green and red curves correspond to zone 1, zone 2 and zone 3, respectively

## 3.2 Measurements of oxygen profiles in the ZoC device

Figure 4a-i depicts the ZoC device with integrated oxygen sensor sections, while Figures 4a-ii and 4a-iii show the culture chamber with these sections visualized using fluorescence microscopy. Initially, the entire device was stabilized at 19% oxygen, indicated by the green fluorescence of the sensors across all regions as shown in the figure 4a-ii. At the end of the experiment, the oxygen scavenger sodium sulfite was introduced into the culture chamber, causing a global decline in oxygen concentration to 0%. This change is observable through a color shift in the sensor sections, turning them orange (Fig 4-iii). Figure 4b provides an additional bright field image of the culture chamber

during the experiment, showcasing cells scattered on the culture surface alongside a portion of a sensor section beneath, with the further underlying parallel gas channels visible in the background.

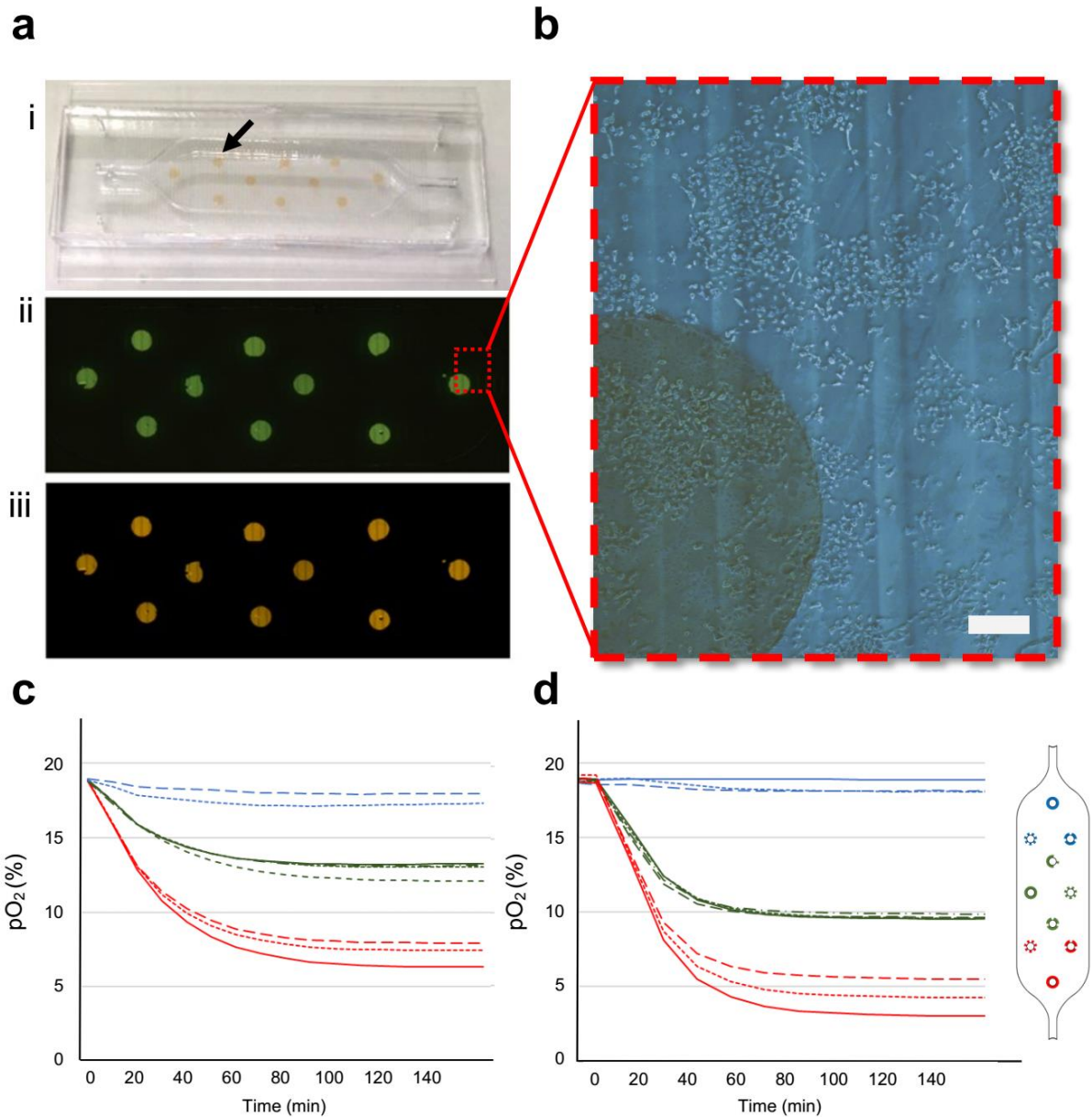

**Fig. 4** Oxygen measurement experiment conducted on the ZoC. **a-i** Displays the device equipped with ten distributed sensor sections across the culture area prior to measurements. An arrow indicates one of the sensors in the culture chamber. **a-ii** Shows fluorescence microscopy images of the sensors in the device at 19% percent oxygen, where the PtTFPP dye is strongly quenched, making the sensors appear green. In contrast, **a-iii** shows the sensors at 0% oxygen after the introduction of the sodium sulfate oxygen scavenger solution, allowing the PtTFPP dyes to emit light without quenching, resulting in the sensor sections appearing as orange, and providing a calibration point at the conclusion of the experiment. **b** Bright field image of the culture chamber with a sensor section during the experiment, showcasing cells scattered on the culture surface, with the parallel lines of gas channels visible in the background. The scale bar is 100 μm. The dynamics of oxygen concentration are shown through the readouts of each sensor sections in the device without cells **c** and with cells in the device **d**, respectively

To support the simulation results, ratiometric oxygen concentration measurements were conducted using sensor sections under two scenarios: with and without cells. The results of these measurement are presented in Figures 4c and 4d, respectively. In both scenarios, all three zones reached distinct oxygen concentration equilibrium. zone 1 maintained a normoxic condition at 19%, consistent with the concentration in the cell incubator, whereas zones 2 and 3 reached lower equilibrium levels. Notably, in zone 2, the oxygen concentration was approximately 12-14% in the absence of cells and 10-11% with cells present. In zone3, there was a similar difference between the scenarios, with concentrations ranging from 6-8% in cell-free experiment and 3-5.5% with the cells in the device.

### 3.3 Phenotype response to the ZoC oxygen gradient

To evaluate the viability of on-chip cultured cells throughout the experiment, we conducted live/dead analysis. The analysis was performed 24 hours after initiating the gas flow in the gas channels under a flow condition of 0.5 µL/min (26 hours post cell seeding). Representative fluorescence microscopy images of zone 3 are shown, with panel (i) displaying live cells in green and panel (ii) showing all cell nuclei in blue, with dead cells highlighted in red, demonstrating high viability (Figure 5a). Remarkably, over 95% of cells maintained viability across all three zones, indicating robust cell survival under these culture conditions (Figure 5b). Cell viability is illustrated for each zone, with zone 1 exhibiting the highest oxygen levels and zone 3 the lowest.

Albumin expression was detected through on-chip immunofluorescent staining in the cells, followed by imaging and quantification. Representative fluorescence microscopy images from the three zones display albumin expression (red) and cell nuclei (blue) within the device after the 26-hour experiment (Figure 5c). Albumin production by the cells in all three zones was quantified based on the scanned imaging sites, as depicted in Figure 5d. Each data point in Figure 5d corresponds to an imaging position within the cell culture chamber, with 36 imaging positions allocated per zone. The albumin production is highest in the periportal region (zone 1) and lowest in the pericentral region (zone 3), corresponding to the highest and lowest oxygen tension levels, respectively. Albumin production in the intermediate region (zone 2) falls between these two. The cumulative data for each zone, demonstrate significantly different albumin expression levels that correlate with the oxygen concentrations (Figure 5e).

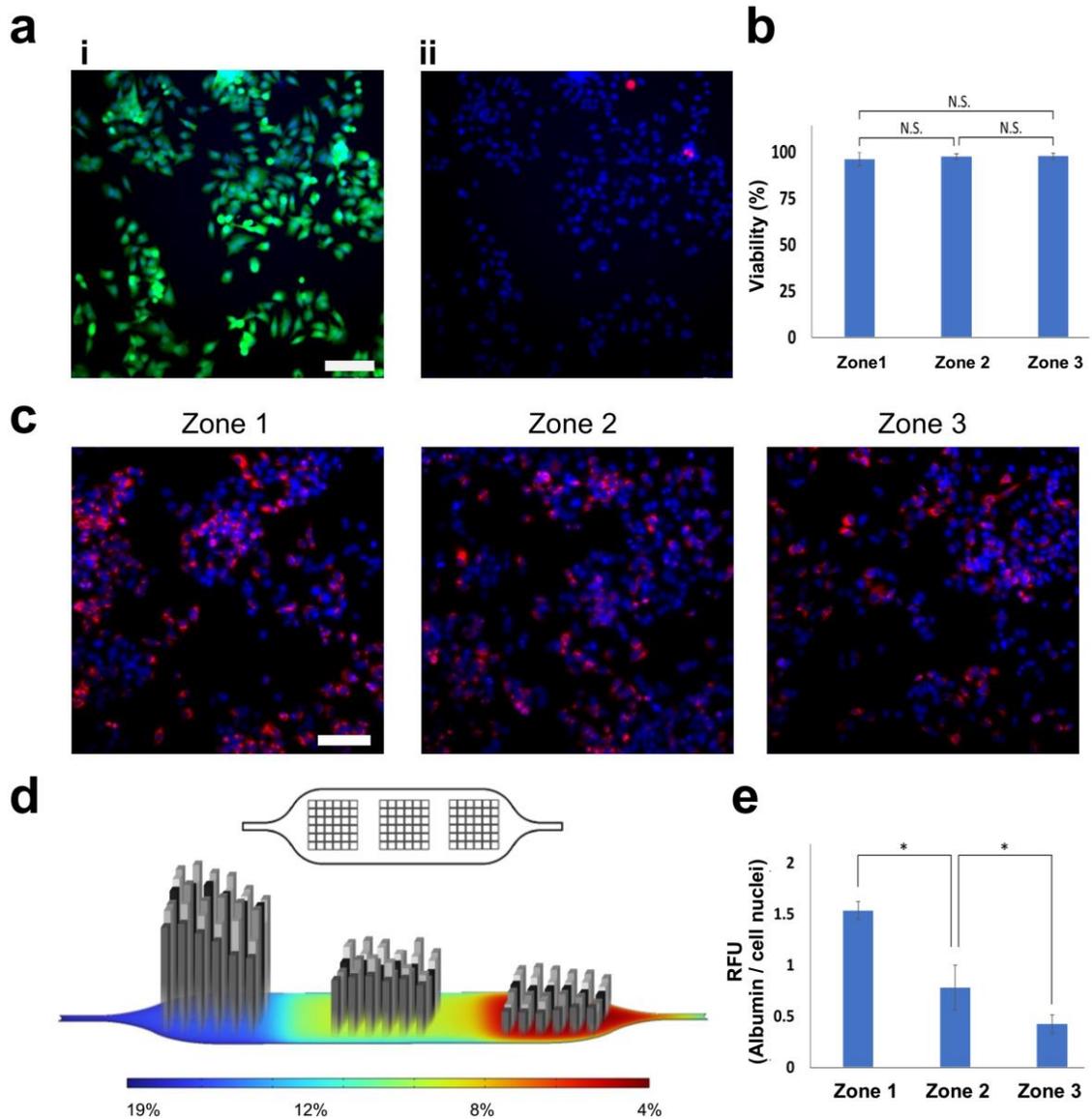

**Fig. 5** Quantitative assessment of HepG2 cell culture viability and functionality within the ZoC device. **a** Fluorescence microscopy images acquired from zone 3 show cells stained with the cell nuclei dye (Hoechst, blue) and the cytosolic live cell marker (calcein-AM, green) **i**, as well as the dead cell marker (ethidium homodimer-1, red) **ii**. **b** Presents the quantitative viability data for the three zones from three separate experiments. **c** Visualization of synthesized intracellular albumin (in red) and cell nuclei (in blue) using fluorescence microscopy, with representative images shown from each zone. Quantification of intracellular albumin levels involves measuring the red signal intensity normalized to the blue signal, presented as intensity per image within each zone **d** or as the average result per zone **e**. Data is acquired from three separate experiments. The scale bar in both the live-dead and the albumin immunostaining images is 100 μm

## 4 Discussion

We have engineered a microfluidic-based *in-vitro* cell culture device, termed the ZoC device, to replicate hepatic zonation. Initially, we designed, simulated, and fabricated a user-friendly device capable of generating and maintaining an oxygen gradient within a cell culture chamber. Next, we integrated oxygen sensors into the device to monitor the generated gradients and validate the simulated results. Finally, we verified the formation of zonation in cultured cells within the device by quantifying albumin production, which exhibited correlation with oxygen concentration, similar to the behavior observed in hepatocytes within sinusoids *in-vivo*.

Tissue oxygen concentration is influenced by various factors such as metabolism, local blood flow, and tissue thickness (Secomb, 2017). For instance, in the liver acinus, oxygen levels vary significantly, ranging from 8.4% to 9.1% in the periportal area to as low as 4.2% to 5.0% in the pericentral zone (Kietzmann, 2019). Traditional cell culturing approaches fail to replicate the structural complexities of the liver, resulting in uniform oxygen distribution, typically dictated by the incubator environment. Furthermore, routine handling procedures of cell cultures such as medium exchange can disrupt oxygenation (Place et al., 2017), leading to limited control over oxygen levels during culture. Although oxygen is a key driver of hepatocyte zonation, nutrient gradients (including but not limited to glucose, hormones and cytokines) also contribute to this phenomenon (Kietzmann, 2017). In conventional cultures without perfusion, nutrients are constantly depleted by the cells and are only replenished when the media is changed. This leads to fluctuations in nutrient and oxygen availability, which undermines their reliability and significance. Microfluidic technology offers a solution by providing precise control over the cellular microenvironment, allowing custom regulation of oxygen levels.

In our microfluidic-based ZoC device, the dual-layer design and the high gas permeability of PDMS facilitate oxygen diffusion from the gas channel through the overlaying PDMS layer into the cell chamber, creating three distinct oxygen concentration zones that mimic *in-vivo* zonation. This setup allows for experiments using the same cell culture batch within the same device, but exposing cells to different oxygen levels, thereby enhancing significance and reducing inter-experimental variations. In addition, conducting parallel culture experiments under varied oxygenation conditions would be inefficient in terms of time and resources. Furthermore, the possible paracrine effects of cells in different zones are completely overlooked when studies are conducted in separate culture flasks or dishes. Conversely, our ZoC device provides three distinct oxygen concentrations within a single device, along with controlled shear stress, utilizing nitrogen and ambient air to achieve desired oxygen levels. As cells are exposed to decreasing oxygen and nutrient levels sequentially along the culture chamber length, downstream cultured cells detect metabolites secreted by upstream cells. Therefore, the device is well-suited for studying cell communication between specialized cells.

Establishing an optimal balance between a physiologically relevant glucose distribution and minimal shear stress while effectively supplying oxygen across the three device zones is essential. Experimental parameters were optimized using CFD simulations prior to device fabrication. We selected a flow rate of 0.5 μL/min to maintain low shear stress (around $2.8 \times 10^{-6}$ Pa), aligning with the suggested range for enhanced hepatic cell functionality (Rashidi et al., 2016). While higher flow rates also maintain suitable shear stress (Figure S1) (Tanaka, Yamato, Okano, Kitamori, & Sato, 2006; Tilles, Baskaran, Roy, Yarmush, & Toner, 2001), our choice aimed to generate a glucose gradient reflecting fasting glucose concentration in liver sinusoids (4.4-6.7 mol/m$^3$ (Edgerton et al., 2006)), evident in our ZoC. Our CFD results demonstrate a 24% decrease in glucose concentration along the ZoC flow direction, from 5.5 to 4.2 mM, emphasizing how altering flow rates can affect nutrient gradients (Figure S2). The nutrient gradient formation is influenced by cell type and metabolic rate, allowing our device to introduce media at various flow rates for cell type-specific glucose gradients. (For a detailed discussion on using the CFD analysis on organ-on-a-chip systems please refer to our previous study (Mahdavi, Hashemi-Najafabadi, Ghiass, & Adiels, 2024).)

Our simulation model assumes PDMS saturation with oxygen at the high-oxygen gas channel due to the significantly higher gas flow rate exceeding saturation thresholds, an approach that conserve computational resources (for calculation, see the SI). We adopt a partition coefficient of 10 for oxygen transfer between culture medium and PDMS, as per previous work (Shiku et al., 2006). However, experimental observations indicate slower equilibration times up to two hours, possibly due to the interface effect or protein build-up on chamber walls, not included in our model assumptions. To address potential travel time in the gas tubes, we conducted dedicated experiments mirroring oxygen measurements, refining the simulation model (SI video 1 and Figure S3). Furthermore, in our actual oxygen measurements, the sensors are positioned beneath a thin, spin-coated PDMS layer of approximately 10 μm, which could potentially introduce additional latency. However, given the thinness of the PDMS layer, this effect is considered neglectable. Notably, we did not account for gradual protein sedimentation on the chamber walls in our model assumptions. Over time, protein accumulation on the PDMS surface may reduce its effectiveness in supplying oxygen to the cells, thereby decreasing oxygen transfer at the interface (Zanzotto et al., 2004). This protein build-up primarily impacts long-term PDMS-based microfluidic cell cultures and likely contributes to the slower equilibration observed in our experiments. However, for short-term experiments like ours, this effect is less concerning.

The equilibrium oxygen concentration in each zone is primarily determined by the gases supplied to the gas layer. However, because the device is not sealed against the ambient environment, some gas exchange also occurs between the environment and the cell chamber. This exchange is minor compared to the impact of cellular oxygen consumption. Both simulations and experiments demonstrated that a sufficiently large number of cells can significantly affect the equilibrium oxygen concentration.

On the fabrication side, maintaining a consistent device thickness is essential for achieving reproducible oxygen gradients as thicker devices create stronger diffusion barriers, potentially leading to inconsistencies in oxygen concentrations across different experiments. While creating gaseous insulation around the upper and lateral regions of the device could potentially create an even more regulated oxygen environment (Esch, Mahler, Stokol, & Shuler, 2014; Rafat, Raad, Rowat, & Auguste, 2009; Wasay & Sameoto, 2015), in this study, we intentionally relied on the ZoC device being exposed to ambient air. This design choice of omitting the insulation prioritizes the user-friendliness of the setup. Additionally, the chip allows for easy access to the cells, and the device fabrication requires one less assembly step compared to insulated counterparts. The separation of the gas channels from the culture chamber mitigates contamination risks, allowing the use of non-sterile gas sources without compromising experimental integrity.

For our experimental oxygen measurements, we utilized circular sensor sections scattered across the culture surface as probes. This approach was chosen to avoid significant alterations of oxygen levels within the chip, which would have occurred if we had used an entire sensor film due to the higher oxygen diffusion properties of PDMS compared to PS (sensor matrix). Additionally, using the entire sensor film would have compromised fluorescence cell imaging due to the presence of spectrally overlapping oxygen sensing dyes. The sensor sections were placed at zone boundaries where discrepancies were likely due to proximity to the uninsulated device border or zonal changes, and measurements confirmed a relatively uniform oxygen distribution within the zones with a concentration variation of only 2%-3% within the zones. The most significant concentration variation occurred in the low oxygen region (zone 3), probably because of the largest oxygen concentration difference between the zone 3 and the ambient atmosphere.

During a cell-free experiment, we discovered that prolonged illumination (455 nm) could cause bleaching of sensor sections, impairing their functionality. The issue was identified after one sensor section in a cell-free experiment was accidentally exposed illumination during the entire experiment for a few hours, causing an abnormal oxygen readout of well above 19%. The affected sensor flake from the experiment was hence omitted from the dataset (Figure 4c) which can be seen in full in Figure S3.

Our study demonstrates a high cell viability across all three oxygen zones in the ZoC, validating the minimally invasive approach to oxygen tension modulation within our device, indicating its capability to support cell culture under varying oxygen conditions. The high viability is attributed to the passive diffusion of oxygen through the PDMS material, eliminating the need for neither oxygen suppliers nor scavengers. Remarkably, cells remained highly viable even in zone 3, which had the lowest oxygen level, suggesting the device capability to support cell culture under varying oxygen conditions.

On-chip immunostaining for quantifying albumin production provides distinct advantages over conventional ELISA-based measurements. While ELISA is the gold standard for measuring secreted albumin protein in liver chips (Chen, Miller, & Shuler, 2018; Du et al., 2021; Vernetti et al., 2016; Ya et al., 2021), in our ZoC, it would provide albumin measures from the outlet media, representing only an average of production from the three zones. In contrast, our approach visualizes the intracellular albumin production patterns locally within the three zones. By fixing the cells, staining them for albumin content, and acquiring an array of 36 fluorescence microscopy images for each zone, we achieved high-resolution quantification of local albumin production. This ensured obtaining data from within each distinct oxygen zone rather than from areas lying in the borders between two zones. Here, the positioning of the arrays was guided by the CFD analysis. Ultimately, the achieved three distinct oxygen zones demonstrate successful functionality, which was supported by the CFD analysis.

Even in standard incubators operating under normoxia in 19% oxygen, it has been shown that high-oxygen-consuming cells like hepatocytes can decrease the oxygen concentration to hypoxic levels (Al-Ani et al., 2018; Place et al., 2017). However, we have demonstrated that our chip can maintain normoxic oxygen concentrations in the periportal zone (zone 1) even in the presence of cell culture. This can be attributed to the oxygen supplement from both bottom and top of the culture chamber. Even with the distinct production levels of albumin, confirming the zonation, we observed slightly higher albumin production on one longitudinal side of the device, likely influenced by the gas channel's configuration. This configuration slightly impacts oxygen availability to cells at the chamber's border, thus affecting albumin synthesis (i.e., the increased fluorescent albumin intensities in the corresponding images within each zone).

## 5 Conclusion

In conclusion, this study presents a novel microfluidic device designed to induce zonation in liver hepatocytes by establishing three distinct oxygen tension levels within the cell culture area. By incorporating a gradient of nutrients and oxygen, this device effectively mimics the physiological conditions of the liver microenvironment. Through simulations and ratiometric oxygen sensing, we validated the device's capability to establish an oxygen gradient along its length.

Quantification of albumin production by HepG2 cells using immunostaining and fluorescence microscopy revealed significant differences across the three oxygen zones, confirming the device's ability to elicit metabolic responses corresponding to varying oxygen concentrations. Despite minor discrepancies in oxygen readings due to sensor placement and border effects, the overall trend demonstrates the device effectiveness in modulating oxygen levels and studying cellular responses.

The simplicity and accessibility of the proposed device, which utilizes readily available laboratory equipment such as nitrogen line and miniature vacuum pump, contribute to its versatility. This accessibility not only facilitates zonation research but also extends its utility to studies requiring oxygen gradients or hypoxic environments, eliminating the need for costly oxygen-modulating infrastructure. The relatively rapid attainment of equilibrium oxygen levels within approximately two hours, and the sustained high cell viability throughout the experiment further support the ZoC's suitability for cell culture applications.

Finally, while initially tailored for liver zonation studies, the adaptability of this device suggests potential applications in diverse research areas requiring controlled oxygen tension or gradient environments. The flexibility of this technology opens up new avenues for exploration in fields ranging from tissue engineering to cancer biology, where precise oxygen modulation is critical for understanding cellular behavior and disease progression. This work lays the groundwork for future investigations leveraging the physiological relevance and accessibility of oxygen-modulating microfluidic devices in cell biology research.